\documentclass{ifacconf}

\makeatletter
\let\old@ssect\@ssect 
\makeatother

\usepackage[bookmarks,bookmarksnumbered,colorlinks]{hyperref}
\hypersetup{linkcolor = black,anchorcolor = black,citecolor =
	black,filecolor = black,urlcolor = black}

\makeatletter
\def\@ssect#1#2#3#4#5#6{%
  \NR@gettitle{#6}
  \old@ssect{#1}{#2}{#3}{#4}{#5}{#6}
}
\makeatother

\usepackage{cancel}
\usepackage{graphicx}      
\usepackage{natbib}        
\usepackage{tikz}
\usepackage{float}
\usepackage{amsfonts}
\usepackage{mathtools}
\usepackage{verbatim}

\usepackage{color}
\definecolor{darkgreen}{rgb}{0,0.5,0.1}

\newtheorem{theorem}{Theorem}
\newtheorem{proposition}[theorem]{Proposition}
\newtheorem{lemma}[theorem]{Lemma}
\newtheorem{corollary}[theorem]{Corollary}
\newtheorem{remark}[theorem]{Remark}
\begin{document}
\begin{frontmatter}

\title{Model order reduction for the TASEP Master equation\thanksref{footnoteinfo}} 

\thanks[footnoteinfo]{This research   is partially supported by the DFG research grants GR 1569/24-1 and KR 1673/7-1,  project number 470999742.}

\author[First]{K. Pioch} 
\author[First]{T. Kriecherbauer} 
\author[Second]{M. Margaliot}
\author[First]{L. Gr\"une} 

\address[First]{University of Bayreuth, Institute of Mathematics, 95440 Bayreuth, Germany (e-mail: kilian.pioch, thomas.kriecherbauer, lars.gruene@uni-bayreuth.de)}
\address[Second]{Tel Aviv University, School of Electrical Engineering, Tel Aviv 6997801, Israel (e.mail: michaelm@tauex.tau.ac.il)}

\begin{abstract} 
The totally asymmetric simple exclusion process (TASEP) is a stochastic model for the unidirectional dynamics of interacting particles on a $1$D-lattice 
that is much used in systems biology and statistical physics. Its master equation describes the evolution of the probability distribution on the state space. The size of the master equation
grows exponentially with the length of the lattice. It is known that the complexity of the system may be reduced using mean field approximations. We provide a rigorous derivation and a stochastic interpretation of these approximations and present numerical results on their accuracy for a number of relevant cases.
\\[-5mm]
\end{abstract}


\end{frontmatter}

\section{Introduction}
The \emph{totally asymmetric simple exclusion process}~(TASEP) 
is a central model in nonequilibrium  statistical mechanics. 
It was first introduced to model
the flow 
of    ribosomes along the mRNA strand during
translation~\citep{MacDonald1968KineticsOB}.
TASEP describes unidirectional dynamics of particles on a 1D chain
where motion is triggered by a stochastic process. Each site is either empty or contains a single particle. 
This exclusion condition indirectly couples the motion of different
particles, as no particle can move to a site that is already occupied.
TASEP has been used to model numerous
natural and artificial processes including vehicular traffic, the kinetics of molecular motors, and
ribosome flow along the mRNA during translation, see e.g.~\cite{TASEP_REV_SHARMA,TASEP_tutorial_2011,ScCN11} and references therein. More on the theoretical side, TASEP is arguably the best studied model for stochastic dynamics in the Kardar-Parisi-Zhang~(KPZ) universality class.
One of the numerous exciting features of TASEP that also plays a role in the present paper is that it exhibits phase transitions between low-density, high-density, and maximum-current phases~\citep{Krug1991,Blythe2007}.

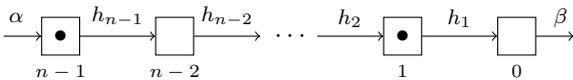
\begin{figure}[htb]
\centering
\begin{tikzpicture}

\node[rectangle,draw,minimum width=0.5cm, minimum height=0.5cm, label=below:\scriptsize{$n-1$}] (e0) at (0,0) {\color{black}$\bullet$};
\node[rectangle,draw,minimum width=0.5cm, minimum height=0.5cm, label=below:\scriptsize{$n-2$}] (e1) at (1.5,0) {\color{black}$\,$};
\node[rectangle,draw,minimum width=0.5cm, minimum height=0.5cm, label=below:\scriptsize{$1$}] (e2) at (4.5,0) {\color{black}$\bullet$};
\node[rectangle,draw,minimum width=0.5cm, minimum height=0.5cm, label=below:\scriptsize{$0$}] (e3) at (6,0) {\color{black}$\,$};
\node[rectangle] (e4) at(3.05,0){\color{black}$\cdots$};

\draw [->] (-0.75,0) to node[above=7, left=-4]{\small{$\alpha$}}(e0);
\draw [->] (e0) to node[above]{\small{$h_{n-1}$}}(e1);
\draw [->] (e1) to node[above]{\small{$h_{n-2}$}}(2.625,0);
\draw [->] (3.375,0) to node[above]{\small{$h_{2}$}}(e2);
\draw [->] (e2) to node[above]{\small{$h_1$}}(e3);
\draw [->] (e3) to node[above=7, right=-4]{\small{$\beta$}}(6.75,0);

\end{tikzpicture}
\caption{Schematic description of TASEP. Particles (circles) hop unidirectionally between~$n$ sites (squares).}
\label{3_node_network}
\end{figure}

In this paper we consider finite one-dimensional lattices with open boundaries as shown in Figure~\ref{3_node_network}.
They consist of $n$ sites labeled in decreasing order from $n-1$ to $0$. Each site is either empty or occupied by exactly one particle. Particles may only move between neighboring sites from left to right. 
In addition, particles can enter the chain at site~$n-1$ and leave it at site~$0$. Each arrow also represents a Poisson process (clock) and these clocks are all stochastically independent. Their rates may all be different. They are denoted by $\alpha$, $\beta$, and $h_k$ with $1 \leq k \leq n-1$.
If the clock associated with rate~$h_k$ rings then a particle moves from site~$k$ to $k-1$ provided site~$k$ is occupied and site~$k-1$ is empty. If the clock associated with rate~$\alpha$ rings then a particle enters the chain if site~$n-1$ is empty. A particle at site~$0$ leaves the chain if the clock associated with rate~$\beta$ rings. Otherwise, no motion occurs.

The rules just described lead to a Markov process in continuous time on a finite state space $\mathbb{S}=\{0,1\}^n$ that has~$2^n$ elements. A central quantity for this stochastic process is the map $t \mapsto x(t) = \{x_i(t)\}_{i \in \mathbb{S}}$, 
where~$x_i(t) \in[0,1]$ 
denotes the probability of the process being in state~$i$ at time~$t$. It is well known that~$x$ satisfies a linear ODE $\dot{x}=\mathbf{A}x$ which is called the \emph{Kolmogorov equation} or the \emph{master equation}.
\begin{figure}[ht]
{\small 
\centering
\begin{tikzpicture}

\node[rectangle,draw] (e0) at (0,0) {\color{purple}$000$};
\node[rectangle,draw] (e1) at (2,-1) {\color{purple}$100$};
\node[rectangle,draw] (e2) at (2,1) {\color{purple}$001$};
\node[rectangle,draw] (e3) at (4,-1) {\color{purple}$101$};
\node[rectangle,draw] (e4) at (4,1) {\color{purple}$010$};
\node[rectangle,draw] (e5) at (6,-1) {\color{purple}$110$};
\node[rectangle,draw] (e6) at (6,1) {\color{purple}$011$};
\node[rectangle,draw] (e7) at (8,0) {\color{purple}$111$};

\draw [->] (e0) to node[left=2, below]{$\alpha$}(e1);
\draw [->] (e2) to node[left=2, above]{$\beta$}(e0);
\draw [->] (e1) to node[below=10, left=12]{$h_2$}(e4);
\draw [->] (e2) to node[above=12, left=14]{$\alpha$}(e3);
\draw [->] (e3) to node[below]{$\beta$}(e1);
\draw [->] (e4) to node[above]{$h_1$}(e2);
\draw [->] (e3) to node[below=10, left=12]{$h_2$}(e6);
\draw [->] (e4) to node[above=12, left=14]{$\alpha$}(e5);
\draw [->] (e5) to node[below]{$h_1$}(e3);
\draw [->] (e6) to node[above]{$\beta$}(e4);
\draw [->] (e6) to node[right=2, above]{$\alpha$}(e7);
\draw [->] (e7) to node[right=2, below]{$\beta$}(e5);
\draw [->] (e0) to (e1);
\draw [->] (e2) to (e0);
\draw [->] (e1) to (e4);
\draw [->] (e2) to (e3);
\draw [->] (e3) to (e1);
\draw [->] (e4) to (e2);
\draw [->] (e3) to (e6);
\draw [->] (e4) to (e5);
\draw [->] (e5) to (e3);
\draw [->] (e6) to (e4);
\draw [->] (e6) to (e7);
\draw [->] (e7) to (e5);
\end{tikzpicture}\caption{Transition rates between all states  
in TASEP with~$n=3$ sites.}
}
\label{Figure2}
\end{figure}
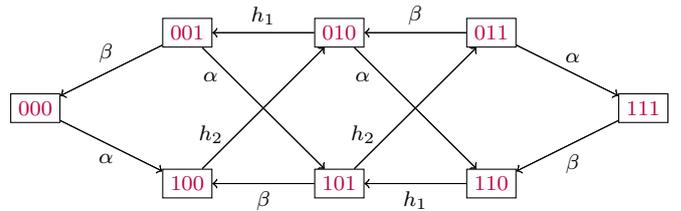
For example, in the case of a lattice with three sites, the master equation can be derived form the transitions between the eight states that are displayed in Figure~2
together with their corresponding rates.
Let us consider state $010$ that denotes the state where only the site in the middle is occupied. This state has
two outgoing and two incoming arrows. The outgoing ones denote a possible evolution from state $010$ into states $110$ and $001$ with rates $\alpha$ and $h_1$, respectively. Similarly, states $011$ and $100$ can evolve into state $010$ with rates $\beta$ and $h_2$, respectively. Using the decimal representation of binary numbers, $010_2=2$, $011_2=3$, $100_2=4$ we arrive at the following expression for the derivative of~$x_{010}\equiv x_2$:
\begin{equation}\label{row3meq}
\dot{x}_2 = -(\alpha + h_1) x_2 + \beta x_3 + h_2 x_4.
\end{equation}
Observe that this corresponds to the third component of the master equation $\dot{x}=\mathbf{A}x$ with $x=(x_0, \ldots, x_7)$. 

The state space for lattices with $n$~sites has size $2^n$. Thus the dimension of the matrix $\mathbf{A}$ 
grows exponentially. Despite being a sparse matrix, $\mathbf{A}$ is irreducible and there are no trivial reductions to lower dimensions, as every state can evolve into any other state with positive probability. 
In some applications, e.g.~modeling mRNA translation,
the chain may consist of hundreds of sites ruling out any numerical simulations of the master equation. However, often 
 one is interested in quantities that are far less detailed than the probability distribution on the full state space. Typical examples for this are the expected occupations $\langle \tau_k(t) \rangle$ of the sites~$k$, $0 \leq k \leq n-1$. Here $\tau_k(t)$ denotes the random variable that takes the value $1$ if site~$k$ is occupied at time $t$ and the value $0$, otherwise. The brackets $\langle \cdot \rangle$ stand for taking the expected value with respect to the probability measure $\mathbb{P}$ associated with the Markov process on the probability space $(\Omega,\mathcal{A},\mathbb{P})$.

As we recall in Section~\ref{Sec:2} below one can derive expressions for $\frac{d}{dt} \langle\tau_k\rangle$ from the master equation, see Proposition~\ref{Prop:exact_first_order} where $\xi_k := 1-\tau_k$. However, they do not form a {\em closed} system of~ODEs, as the right-hand side contains the unknown $2$-point correlations $\langle \tau_k \xi_{k-1}\rangle = \langle \tau_k \rangle-\langle \tau_k \tau_{k-1} \rangle$. One may close the system by using the approximation $\langle \tau_k \tau_{k-1} \rangle \approx \langle \tau_k \rangle \langle \tau_{k-1} \rangle$.
This leads to the well-known ribosome flow model (RFM), a system of non-linear ODEs with~$n$ 
rather than $2^n$ states 
that 
has been analyzed using tools from systems and control theory, see e.g.~\cite{Ofir2023-tx,Margaliot2012-nj,Raveh2016}.

In many cases the solutions of the RFM yield an excellent approximation of $\{\langle\tau_k(t)\rangle\}_{0\leq k\leq n-1}$.
However, it is also known that there are cases with significant deviations between them, see e.g.~\citep{Blythe2007}. 
In order to reduce these deviations, in this paper 
\begin{itemize} \item we develop a systematic approach for constructing a hierarchy of ODE systems containing higher-order approximations of correlations,
\item provide an expression of the error that can be used for identifying situations in which higher-order models provide better approximations,
\item and evaluate the resulting errors by a thorough numerical analysis\footnote{The software developed for performing our simulations can be found at https://github.com/KilianPiochUBT/TASEP.}.
\end{itemize}

The idea to use approximations of higher-order correlations in order to arrive at a closed ODE system is not new. It can be seen as a as special case of a {\em Moment Closure}, see~\citep{Kuehn2016} for a recent review. In the classification of this review the method we use here falls into the class of {\em microscopic closures}. These have been used, e.g., in the context of epidemiological models \citep{Kiss2017-cw}, where in some cases they yield exact rather than merely approximating closures, or to random sequential absorption \citep{PhysRevA.45.8358}, where the idea of overlapping approximations for arbitrary order $m$, see~\eqref{eq:overlap_approx}, below, already appears. In the context of TASEP, such approximations were used in \cite{Pelizzola2017-ox} for order $m=2$ and $3$. In this paper, building on the derivation of the ODE system explained in \cite{DerE97} (see also \cite{Blythe2007}), we provide a complete description of this system for arbitrary chain length in Theorem \ref{thm:model} and define closures with approximations of arbitrarily high order $m$ in Remark~\ref{rem:closure}.

In general, as noted in \citep{Kuehn2016}, mathematically rigorous justifications of this approach are rare and usually restricted to particular systems. For TASEP, in Corollary~\ref{cor:approx} we provide a characterization of the error that can be used to identify situations in which such justifications can be derived. We discuss some of these situations in the case study after this corollary and in our numerical results.
 In our error analysis we focus on the equilibria of the ODE systems. This is on the one hand because correctness of the equilibrium provides a suitably simplified setting for the analysis carried out in this paper. On the other hand, as the equilibrium is a global attractor of the master equation, the error at the equilibrium determines the long-time error of the model. Moreover, it provides an interesting connection of the observed error to the different phases in TASEP, shown in the phase diagram in Figure~\ref{fig:phase_diagram}.

The remainder of the paper is structured as follows. In Section~\ref{Sec:2} we derive the full ODE system for the derivatives of all correlations. In Section~\ref{Sec:3} we explain the approximating closures that we use in order to reduce the order of the system from Section~\ref{Sec:2} and analyze the resulting errors for particular settings. In Section~\ref{Sec:4} we provide an evaluation of the error based on numerical simulations. Section~\ref{Sec:5} closes the paper.

\section{Derivatives of $\ell$-point correlations}\label{Sec:2}

\begin{figure}[ht]
\centering
\begin{tikzpicture}
\node[rectangle,draw,minimum width=0.7cm, minimum height=0.7cm, label=below:\tiny{$n-1$}] (e0) at (0,0) {\color{black}$*$};
\node[rectangle,minimum width=0.7cm, minimum height=0.7cm] (e1) at (0.7,0) {\color{black}$\cdots$};
\node[rectangle,draw,minimum width=0.7cm, minimum height=0.7cm, label=below:\tiny{$k+2$}] (e2) at (1.4,0) {\color{black}$*$};
\node[rectangle,draw,minimum width=0.7cm, minimum height=0.7cm, label=below:\tiny{$k+1$}] (e3) at (2.1,0) {\color{black}$\bullet$};
\node[rectangle,draw,minimum width=0.7cm, minimum height=0.7cm, label=below:\tiny{$k$}] (e4) at (2.8,0) {\color{black}$\,$};
\node[rectangle,draw,minimum width=0.7cm, minimum height=0.7cm, label=below:\tiny{$k-1$}] (e5) at (3.5,0) {\color{black}$*$};
\node[rectangle,minimum width=0.7cm, minimum height=0.7cm] (e6) at (4.55,0) {\color{black}$\cdots$};
\node[rectangle,draw,minimum width=0.7cm, minimum height=0.7cm, label=below:\tiny{$0$}] (e7) at (5.6,0) {\color{black}$*$};

\node[rectangle,draw,minimum width=0.7cm, minimum height=0.7cm, label=below:\tiny{$n-1$}] (e8) at (0,-1.2) {\color{black}$*$};
\node[rectangle,minimum width=0.7cm, minimum height=0.7cm] (e9) at (0.7,-1.2) {\color{black}$\cdots$};
\node[rectangle,draw,minimum width=0.7cm, minimum height=0.7cm, label=below:\tiny{$k+2$}] (e10) at (1.4,-1.2) {\color{black}$*$};
\node[rectangle,draw,minimum width=0.7cm, minimum height=0.7cm, label=below:\tiny{$k+1$}] (e11) at (2.1,-1.2) {\color{black}$\,$};
\node[rectangle,draw,minimum width=0.7cm, minimum height=0.7cm, label=below:\tiny{$k$}] (e12) at (2.8,-1.2) {\color{black}$\bullet$};
\node[rectangle,draw,minimum width=0.7cm, minimum height=0.7cm, label=below:\tiny{$k-1$}] (e13) at (3.5,-1.2) {\color{black}$*$};
\node[rectangle,minimum width=0.7cm, minimum height=0.7cm] (e14) at (4.55,-1.2) {\color{black}$\cdots$};
\node[rectangle,draw,minimum width=0.7cm, minimum height=0.7cm, label=below:\tiny{$0$}] (e15) at (5.6,-1.2) {\color{black}$*$};

\draw [->] (e7) to (6.3, 0);

\node[rectangle,draw,minimum width=0.7cm, minimum height=0.7cm, label=below:\tiny{$n-1$}] (e100) at (0,-3) {\color{black}$*$};
\node[rectangle,minimum width=0.7cm, minimum height=0.7cm] (e101) at (1.05,-3) {\color{black}$\cdots$};
\node[rectangle,draw,minimum width=0.7cm, minimum height=0.7cm, label=below:\tiny{$k+1$}] (e102) at (2.1,-3) {\color{black}$*$};
\node[rectangle,draw,minimum width=0.7cm, minimum height=0.7cm, label=below:\tiny{$k$}] (e103) at (2.8,-3) {\color{black}$\bullet$};
\node[rectangle,draw,minimum width=0.7cm, minimum height=0.7cm, label=below:\tiny{$k-1$}] (e104) at (3.5,-3) {\color{black}$\,$};
\node[rectangle,draw,minimum width=0.7cm, minimum height=0.7cm, label=below:\tiny{$k-2$}] (e105) at (4.2,-3) {\color{black}$*$};
\node[rectangle,minimum width=0.7cm, minimum height=0.7cm] (e106) at (4.9,-3) {\color{black}$\cdots$};
\node[rectangle,draw,minimum width=0.7cm, minimum height=0.7cm, label=below:\tiny{$0$}] (e107) at (5.6,-3) {\color{black}$*$};

\node[rectangle,draw,minimum width=0.7cm, minimum height=0.7cm, label=below:\tiny{$n-1$}] (e108) at (0,-4.2) {\color{black}$*$};
\node[rectangle,minimum width=0.7cm, minimum height=0.7cm] (e109) at (1.05,-4.2) {\color{black}$\cdots$};
\node[rectangle,draw,minimum width=0.7cm, minimum height=0.7cm, label=below:\tiny{$k+1$}] (e110) at (2.1,-4.2) {\color{black}$*$};
\node[rectangle,draw,minimum width=0.7cm, minimum height=0.7cm, label=below:\tiny{$k$}] (e111) at (2.8,-4.2) {\color{black}$\,$};
\node[rectangle,draw,minimum width=0.7cm, minimum height=0.7cm, label=below:\tiny{$k-1$}] (e112) at (3.5,-4.2) {\color{black}$\bullet$};
\node[rectangle,draw,minimum width=0.7cm, minimum height=0.7cm, label=below:\tiny{$k-2$}] (e113) at (4.2,-4.2) {\color{black}$*$};
\node[rectangle,minimum width=0.7cm, minimum height=0.7cm] (e114) at (4.9,-4.2) {\color{black}$\cdots$};
\node[rectangle,draw,minimum width=0.7cm, minimum height=0.7cm, label=below:\tiny{$0$}] (e115) at (5.6,-4.2) {\color{black}$*$};

\draw [->] (e107) to (6.3, -3);

\end{tikzpicture}
\caption{Schematics for entering (top rows) or leaving (bottom rows) the set $T_k$. The occupation status of sites marked with $*$ are irrelevant for these dynamics.}
\label{jump_cond}
\end{figure}
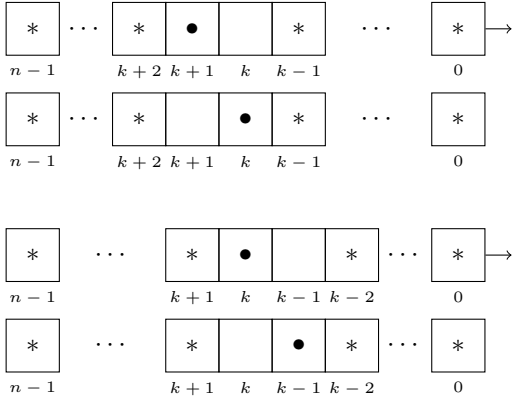

We begin by determining the derivatives $\frac{d}{dt} \langle\tau_k\rangle$. As the random variable $\tau_k(t)$ only takes the values $0$ and $1$, we have~$\langle \tau_k(t) \rangle=\mathbb{P}(\tau_k(t)=1)$. Thus $\langle \tau_k(t) \rangle$ can be expressed as the sum of $2^{n-1}$ components $x_i(t)$ of the probability distribution $x(t)$,
\begin{equation}\label{formula_expectation_tau}
\langle \tau_k(t) \rangle = \sum_{i \in T_k} x_i(t), 
\end{equation}
where $T_k:= \{i \in \mathbb{S} \mid$ site~$k$ is occupied$\}$.
Hence we may use the master equation to determine $\frac{d}{dt} \langle\tau_k\rangle$. Observe first that all dynamics $i \rightarrow j$ within the set $T_k$ do not change the value of $\langle \tau_k(t) \rangle$ because its contributions to~$\dot{x}_i$ and~$\dot{x}_j$ are both given by rate$(i \rightarrow j)$ multiplied by~$ x_i(t)$,  
but with opposite signs so that they cancel in the sum. Clearly, dynamics within the 
complement~$\Xi_k:= \mathbb{S} \setminus T_k$ do not change the value of~$\langle \tau_k(t) \rangle$ either. It remains to consider
all configurations that allow to leave or to enter the set $T_k$. For the cases $1 \leq k \leq n-2$ they are displayed in Figure~\ref{jump_cond}.

Since the transition depicted in the top rows of Figure~\ref{jump_cond} occurs with rate   $h_{k+1}$ and the transition depicted in the bottom rows occurs with rate $h_{k}$, we have
\begin{equation}\label{dot_tau_k}
\dfrac{d}{dt}\langle\tau_k(t) \rangle = h_{k+1} \!\!\!\!\sum_{i \in T_{k+1}\cap\, \Xi_{k}} \!\!\!\! x_i(t) \;\;\;- \;\;\;h_k \!\!\!\!\sum_{i \in T_{k}\cap\, \Xi_{k-1}} \!\!\!\! x_i(t).
\end{equation}
The first sum on the right hand side of \eqref{dot_tau_k} equals the probability that site~$k+1$ is occupied and site~$k$ is empty and can be expressed as an expected value 
\begin{equation}\label{backto_expectations}
\sum_{i \in T_{k+1}\cap\, \Xi_{k}} \!\!\!\! x_i = \langle \tau_{k+1} (1-\tau_k) \rangle = \langle \tau_{k+1} \xi_k\rangle,
\end{equation}
with the notation $\xi_k := 1 - \tau_k$ for the random variable that takes the value $1$ if site~$k$ is empty and the value $0$,
otherwise. 
Extending this argument also for the sites at the boundary we have derived the following
\begin{proposition}\label{Prop:exact_first_order} For $n \geq 2$ and $0 \leq k \leq n-1$ the derivative of $\langle\tau_k(t) \rangle$ exists and satisfies the relation
\begin{equation*}
\begin{aligned}
\dfrac{d}{dt}\langle \tau_0 \rangle &= h_1 \langle \tau_1 \xi_0 \rangle - \beta \langle \tau_0 \rangle,\\
\dfrac{d}{dt}\langle\tau_k\rangle &= h_{k+1} \langle \tau_{k+1} \xi_k\rangle - h_k \langle \tau_k \xi_{k-1} \rangle, \;\; 1\leq k\leq n-2,\\
\dfrac{d}{dt}\langle \tau_{n-1} \rangle &= \alpha \langle \xi_{n-1} \rangle - h_{n-1} \langle \tau_{n-1} \xi_{n-2}\rangle,
\end{aligned}
\end{equation*}
where we have omitted the $t$-dependence for the sake of readability. 
\end{proposition}

Using the ideas just described, we may also derive expressions for the derivatives of $\langle \tau_k \xi_{k-1} \rangle$ that appear on the right-hand side of the equations in Proposition~\ref{Prop:exact_first_order}.
E.g., for $2 \leq k \leq n-2$, one can enter the set $T_{k}\cap\, \Xi_{k-1}$ from $T_{k+1}\cap\, \Xi_{k}\cap\, \Xi_{k-1}$ with rate $h_{k+1}$ and from $T_{k}\cap T_{k-1}\cap\, \Xi_{k-2}$ with rate $h_{k-1}$. The set $T_{k}\cap\, \Xi_{k-1}$ is left only when the clock associated with rate $h_k$ rings. This implies
\begin{equation}\label{exact_second_order}
\begin{aligned}
\dfrac{d}{dt}\langle \tau_k \xi_{k-1} \rangle &= h_{k+1} \langle \tau_{k+1} \xi_k \xi_{k-1} \rangle + h_{k-1} \langle \tau_k \tau_{k-1} \xi_{k-2} \rangle\\
&- h_k\langle \tau_k \xi_{k-1}\rangle.
\end{aligned}
\end{equation}

Again, the right-hand side of equation~\eqref{exact_second_order} contains new terms and more equations need to be added.
In order to derive these equations systematically we 
encode $\ell$-point correlations of the form $\langle X_{d+\ell-1} X_{d+\ell-2} \cdots X_{d+1} X_{d} \rangle$ with $X_k \in \{\tau_k, \xi_k\}$ for $0 \leq d \leq k \leq d+\ell-1 \leq n-1$ by $\ell$-digit binary numbers
\begin{equation}\nonumber
\begin{array}{rccl}
\langle X_{d+\ell-1} X_{d+\ell-2} \cdots X_{d+1} X_{d} \rangle &\equiv& \prescript{\ell}{d}\langle (b_{\ell-1}\cdots b_0)_2 \rangle &= \prescript{\ell}{d}\langle b \rangle \;
\end{array}
\end{equation}
 where, for $0\leq l \leq \ell-1$, we set $b_l = 1$ if $X_{d+l}=\tau_{d+l}$ and $b_l = 0$, otherwise. E.g., 
$$\langle  \xi_7 \xi_6 \tau_5 \tau_4 \xi_3 \rangle \equiv \prescript{5}{3}\langle (00110)_2\rangle = \prescript{5}{3}\langle 6\rangle.$$
 We may write
 $$
 \prescript{\ell}{d}\langle b \rangle = \mathbb{P}(X_{d+\ell-1}\cdots X_{d}=1) =\sum_{i \in Y_{d+\ell-1} \cap \cdots \cap Y_{d}} x_i,
 $$
  where $Y_{d+l}=T_{d+l}$ if $b_l=1$ and $Y_{d+l}=\Xi_{d+l}$, otherwise. We obtain an expression for $\frac{d}{dt} \big(\prescript{\ell}{d}\langle b \rangle\big)$ by determining, for each clock, all of the dynamics it triggers where the set of states $Y_{d+\ell-1} \cap \cdots \cap Y_{d}$
  is left or entered. Let us begin with the leftmost clock that rings with rate $\alpha$, see Figure~\ref{3_node_network}. In this case $\prescript{\ell}{d}\langle b \rangle$ can only change if site~$n-1$ is included in the $\ell$-point correlation, i.e.~if $n-1=d+\ell-1$. Then the set of states associated with $\prescript{\ell}{d}\langle b \rangle$ is left at the ringing of the clock if $b_{\ell-1}=0$ and the chain can move into this set of states if $b_{\ell-1}=1$ (from the set of states associated with $\prescript{\ell}{d}\langle b-2^{\ell-1} \rangle$). 
  In summary, the contribution of the rate-$\alpha$-clock to $\frac{d}{dt} \big(\prescript{\ell}{d}\langle b \rangle\big)$ is given by
\begin{equation}\nonumber
\alpha \left[\prescript{\ell}{d}\langle b - 2^{\ell-1} \rangle\lvert_{\ell+d=n,\,b_{\ell-1}=1}\;- \;  \prescript{\ell}{d}\langle b \rangle\lvert_{\ell+d=n,\,b_{\ell-1}=0} \right].
\end{equation}
Here we have listed the conditions under which the corresponding term appears 
to the right of the vertical line $\vert\,$. Analyzing the effects of each clock one arrives at
\begin{theorem}
For $0\leq d < d+\ell \leq n$ and $0\leq b < 2^\ell$ the derivative of $\prescript{\ell}{d}\langle b \rangle$ exists and satisfies 
\begin{align}
\dfrac{d}{dt} \prescript{\ell}{d}\langle b \rangle &= \alpha \prescript{\ell}{d}\langle b - 2^{\ell-1} \rangle\lvert_{\ell+d=n,\atop b_{\ell-1}=1} \label{al_1} 
+\beta \prescript{\ell}{d}\langle b+1 \rangle\lvert_{d=0,\atop b_0=0}\\
&-\alpha \prescript{\ell}{d}\langle b \rangle\lvert_{\ell+d=n,\,b_{\ell-1}=0} \; -  \; \beta \prescript{\ell}{d}\langle b \rangle\lvert_{d=0, b_0=1}\label{al_2}\\
&+\sum_{j=1}^{\ell-1}\left[h_{j+d}\prescript{\ell}{d}\langle b+2^{j-1} \rangle\lvert_{b_{j} = 0,\,b_{j-1} = 1}\right]\label{al_3}\\
&-\sum_{j=1}^{\ell-1}\left[h_{j+d}\prescript{\ell}{d}\langle b \rangle\lvert_{b_{j} = 1,\, b_{j-1} = 0}\right]\label{al_4}\\
&+h_{\ell+d}\prescript{\ell+1}{d}\langle b + 2^{\ell-1} \rangle\lvert_{\ell+d<n,\,b_{\ell-1} = 1}\label{al_5} \\
&+ h_d \prescript{\ell+1}{d-1}\langle 2(b+1) \rangle\lvert_{d>0,\,b_{0} = 0}\label{al_51}\\
&-h_{\ell+d}\prescript{\ell+1}{d}\langle b+2^\ell \rangle\lvert_{\ell+d<n,\,b_{\ell-1} = 0}\label{al_6} \\
&- h_d \prescript{\ell+1}{d-1}\langle 2b \rangle\lvert_{d>0,\,b_{0} = 1}\label{al_61}\, .
\end{align}
\vspace*{-3mm}
\label{thm:model}
\end{theorem}
{\bf Proof:}
Lines \eqref{al_1}-\eqref{al_2} correspond to the dynamics of particles entering or leaving the chain, \eqref{al_3}-\eqref{al_4} come from the dynamics between sites that are both contained in the range of sites defining the $\ell$-point correlation, and \eqref{al_5}-\eqref{al_61} are due to the dynamics of particles from site~$d+\ell$ to site~$d+\ell-1$ and from site~$d$ to site~$d-1$. For more detais, see~\cite{Pioch23}.
\qed

It follows from the terms in \eqref{al_5}-\eqref{al_61} that the expression for the derivative of an $\ell$-point correlation contains $(\ell+1)$-point correlations if $\ell < n$. Starting with the equations of Proposition~\ref{Prop:exact_first_order} for the $1$-point correlations $\langle \tau_k \rangle$ and extending the system until it is closed
we arrive at a system that contains at least one $n$-point correlation. 
Then, by the irreducibility of the master equation, all of the~$2^n$ different $n$-point correlations must appear in the closed system. Therefore, it is of larger dimension than the master equation and nothing is gained. Clearly, we need approximations to reduce the model order.

\section{Closures via approximations of~correlations}\label{Sec:3}

As described in the survey~\citep{Kuehn2016}, 
a standard method to reduce the complexity of a microscopic model like TASEP is to
restrict the length of the correlations at some order~$m<n$ using pair-approximations. For~$m=1$ this means, e.g., replacing $\langle \tau_k \xi_{k-1} \rangle$ by $\langle \tau_k\rangle \langle \xi_{k-1} \rangle$ which transforms the equations of Proposition~\ref{Prop:exact_first_order} to the ribosome flow model as we explained in the introduction. The following lemma is the basis for our error analysis of the pair-approximation in the context of TASEP.

\begin{lemma}
Suppose $X$, $Y$, and $Z$ are random variables that take values in $\{0, 1\}$ with $\langle Y \rangle > 0$.  Then the equation
\begin{eqnarray*}
&& \langle XYZ \rangle - \frac{\langle XY \rangle \langle YZ \rangle}{\langle Y \rangle}\\
&& \; = \; 
\Big(\langle X Z \,|\, Y=1 \rangle - \langle X \,|\, Y=1 \rangle \langle Z \,|\, Y=1 \rangle\Big) \cdot \langle Y \rangle
\end{eqnarray*}
holds, where $\langle V \,|\, Y=1 \rangle$ denotes the conditional expectation of a random variable $V$ under the condition $Y=1$.\\[-4mm]
\label{lemma:XYZ}
\end{lemma}
{\bf Proof:}
Using $\langle V \rangle = \mathbb{P}(\{V=1\})$ for random variables $V$ taking values in~$\{0,1\}$
we obtain
\begin{eqnarray*}
    && \langle X \,|\, Y=1 \rangle \; = \; \mathbb{P}(\{X=1\,|\,Y=1\}) \\
     && = \;  \frac{\mathbb{P}(\{X=1\} \cap \{Y=1\})}{\mathbb{P}( \{Y=1\})}
    \; = \; \frac{\mathbb{P}(\{XY=1\})}{\mathbb{P}(\{Y=1\})} \; = \; \frac{\langle XY \rangle}{\langle Y \rangle}.
\end{eqnarray*}
Likewise, we have
\[\langle Z \,|\, Y=1  \rangle = \frac{\langle YZ \rangle}{\langle Y \rangle} \quad \mbox{and} \quad 
\langle XZ \,|\, Y=1  \rangle = \frac{\langle XYZ \rangle}{\langle Y \rangle}.
\]
Inserting these expression on the right hand side of the claimed identity proves its correctness.
\qed

Writing the $m+1$-fold product~$X_{d+m} \cdots X_{d}$ as~$XYZ$ with $X=X_{d+m} \cdots X_{d+j+1}$, $Y=X_{d+j} \cdots X_{d+i+1}$, and $Z=X_{d+i} \cdots X_{d}$ for some $0\leq i \leq j < m$ the pair-approximation replaces~$\langle XYZ \rangle$ by~$\langle XY \rangle \langle YZ \rangle / \langle Y \rangle$ that is composed of three correlations of order~$\leq m$. The case~$i=j$ is to be interpreted as $Y=1$ denoting the empty product. Note that Lemma~\ref{lemma:XYZ} provides a representation for the error of such an approximation.

In \citep{Pioch23} different choices for the parameters~$i$ and~$j$ were studied. In all cases considered there is overwhelming numerical evidence that the choice of maximal overlap~$Y$ between~$XY$ and~$YZ$, i.e.~$i=0$ and~$j=m-1$, yields the best results. Therefore we restrict our attention to this choice in the present paper. In this case $X=X_{d+m}$, $Y=X_{d+m-1} \cdots X_{d+1}$, and~$Z=X_{d}$. Lemma~\ref{lemma:XYZ} can then be formulated in the following way.

\begin{corollary}
    Let $X_i$, $i=d,\ldots,d+m$,  be random variables taking values in~$\{0,1\}$ with $\langle X_{d+m-1}\cdots X_{d+1}\rangle>0$. Define
    \[ A:= \{\omega\in\Omega\,|\, X_{d+m-1}=\ldots = X_{d+1}=1\}\]
    and, for any random variable $Y$, denote by $\langle Y\,|\,A\rangle$ the conditional expectation of $Y$ given $A$. 
    Then the equation
    \begin{eqnarray*}
      && \langle X_{d+m} \cdots X_{d}\rangle - \frac{\langle X_{d+m} \cdots X_{d+1}\rangle \langle X_{d+m-1} \cdots X_{d} \rangle}{\langle X_{d+m-1} \cdots X_{d+1} \rangle} \\
     &&\;= \; \big(\langle X_{d+m} X_{d}\,|\, A\rangle - \langle X_{d+m}\,|\, A\rangle\langle X_{d}\,|\, A\rangle\big)\\
      && \qquad\qquad 
       \cdot \langle X_{d+m-1} \cdots X_{d+1}\rangle
    \end{eqnarray*}
    holds. In particular, the approximation error is zero iff~$X_{d+m}$ and~$X_{d}$ are conditionally uncorrelated given~$A$.
\label{cor:approx}\end{corollary}

\begin{remark}
We refer to a closed system of differential equations \eqref{al_1}--\eqref{al_61} in which all correlations of length $m+1$ are replaced by the approximation 
\begin{equation}\label{eq:overlap_approx}
\langle X_{d+m} \cdots X_{d}\rangle \approx \frac{\langle X_{d+m} \cdots X_{d+1}\rangle \langle X_{d+m-1} \cdots X_{d} \rangle}{\langle X_{d+m-1} \cdots X_{d+1} \rangle},
\end{equation}
and all equations for correlations of length $\ge m+1$ are removed as a {\em mean-field approximation of order $m$.}
\label{rem:closure}
\end{remark}

The number of correlations of length $k$ is given by $(n-k)2^k$. Summing the upper bound $n2^k$ for this expression for $k=1,\ldots,m$, we obtain the upper bound $n2^{m+1}$ for the total number of correlations of order $\le m$. For instance, if we use a mean-field approximation of order $m=4$ for a chain length of $n=20$, we reduce the model order from $2^n = 1\,048\,576$ states in the master equation to less than $20 \cdot 2^5 = 640$ states.

The numerical results in \cite{Pioch23} and in Section \ref{Sec:4} 
suggest that the approximation error for the proposed approximate closures, i.e., the product on the right hand side of the identity from Corollary \ref{cor:approx} decreases for increasing $m$. We investigate this fact for two situations with $X_k=\tau_k$, which also shed light on the question which of the two factors of the product is more responsible for this decrease.

The first situation we look at is the TASEP model with transition rates~$\alpha=\beta=0.1$, and all~$h_k=1$. It was already observed in~\citep{Blythe2007} that in this case the equilibrium states of RFM and TASEP differ significantly.
We consider the states at the outflow end of the chain where 
a typical bottleneck situation occurs in which the particles ``queue'' before the exit of the chain. 
Using the equilibrium distribution from the master equation one can readily compute all terms on the right hand side of the identity from Corollary~\ref{cor:approx}. Table~\ref{tab:errors} displays the errors 
in the case of chain length~$n=5$ and~$d=0$.

\begin{table}[htb]
    \centering
    \begin{tabular}{c|l|l|l|l|l}
         $m$ &  $a_m$ &  $b_m$ & $c_m$ & $a_mb_m$ & $a_mb_m/c_m$\\ \hline
          1  & 0.0771 & 1 & 0.552 & 0.0771 & 0.140 \\
          2  & 0.0120 & 0.628 & 0.381 & 0.00756 & 0.0198\\
          3  & 0.00146 & 0.424 & 0.238 & 0.000619 & 0.00260 \\
          4 & 0.000162 & 0.265 & 0.119 & 0.0000428 & 0.000360
    \end{tabular}
    \caption{Values for\\ $a_m := \langle \tau_{m} \tau_{0}\,|\, A\rangle - \langle \tau_{m}\,|\, A\rangle\langle \tau_{0}\,|\, A\rangle$,\\
    $b_m := \langle \tau_{m-1} \cdots \tau_{1}\rangle$, $c_m := \langle \tau_{m} \cdots \tau_{0}\rangle$,\\ the approximation error $a_mb_m$ from Corollary \ref{cor:approx} and the relative error $a_mb_m/c_m$, all for TASEP with $\alpha=\beta=0.1$, $h_k=1$ and $n=5$.}
    \label{tab:errors}
\end{table}

The table shows that on the one hand the absolute and the relative approximation errors~$a_mb_m$ and $a_mb_m/c_m$ decrease rapidly when using approximations with higher~$m$ and on the other hand that the first factor $a_m 
$ 
decreases much faster than the second factor $b_m 
$. This suggests that in this example the fact that the error term for higher order approximations involves conditional expectations of~$\tau_m$ and~$\tau_0$ that are further apart from each other is decisive for the decrease of the error.

The second situation we look at is in some sense the opposite of the first. Suppose that we have a triplet of sites $k+2$, $k+1$, and $k$, in which the jump rate from $k+1$ to $k$ is much higher than that from $k+2$ to $k+1$ and from $k$ to $k-1$. We could call this an ``anti-bottleneck'' or a ``fast lane''. Its presence implies that a state of the form $\tau_{k+2}\tau_{k+1}\tau_k = *10_2$ 
is very unlikely, because it will transition into $*01_2$ very quickly. 
As a consequence, $\tau_{k+1}=1$ implies $\tau_k=1$ with high probability, implying a strong correlation between~$\tau_{k+1}=1$ 
and~$\tau_k=1$ and thus the approximation error with~$d=k$ and~$m=1$, i.e., 
\[ \langle \tau_{k+1}\tau_k\rangle - \langle \tau_{k+1}\rangle \langle \tau_{k}\rangle \]
will be large. The situation improves if we increase $m=1$ to $m=2$: Among the $4$ different states satisfying $A$, i.e., $010_2$, $011_2$, $110_2$, and $111_2$, the second and the fourth occur with much higher probability than the first and the third. If we assume that the conditional probabilities of the first and the third equal $\varepsilon_1\approx 0$ and $\varepsilon_3\approx 0$, respectively, this implies $\langle \tau_k\,|\, A \rangle = 1- \varepsilon_1-\varepsilon_3$ and 
\begin{eqnarray*} 
\langle \tau_{k+2}\,|\, A \rangle & = & \underbrace{\mathbb{P}(\tau_{k+2}\tau_{k+1}\tau_k = 110_2 \,|\, A)}_{= \varepsilon_3}\\ && \; + \; \underbrace{\mathbb{P}(\tau_{k+2}\tau_{k+1}\tau_k = 111_2 \,|\, A)}_{=\langle \tau_{k+2}\tau_k\,|\, A \rangle} \\ & = & \langle \tau_{k+2}\tau_k\,|\, A \rangle + \varepsilon_3.
\end{eqnarray*}
Together this gives 
\begin{eqnarray*}
     && \langle \tau_{k+2}\,|\, A \rangle\langle \tau_k\,|\, A \rangle  \; = \;  (\langle \tau_{k+2}\tau_k\,|\, A \rangle + \varepsilon_3)(1- \varepsilon_1-\varepsilon_3)\\
     && \; = \; \langle \tau_{k+2}\tau_k\,|\, A \rangle - \langle \tau_{k+2}\tau_k\,|\, A \rangle(\varepsilon_1+\varepsilon_3) +  \varepsilon_3(1- \varepsilon_1-\varepsilon_3)
\end{eqnarray*}
and thus 
\[ |\langle \tau_{k+2}\tau_k\,|\, A \rangle - \langle \tau_{k+2}\,|\, A \rangle\langle \tau_k\,|\, A \rangle| \le \varepsilon_1 + 2\varepsilon_3.\]

Hence, if $\varepsilon_1$ and $\varepsilon_3$ are close to $0$, then for $m=2$ the approximation error will be small.

\section{Numerical results}\label{Sec:4}

\begin{figure}[ht]
    \centering
    \includegraphics[scale=0.15]{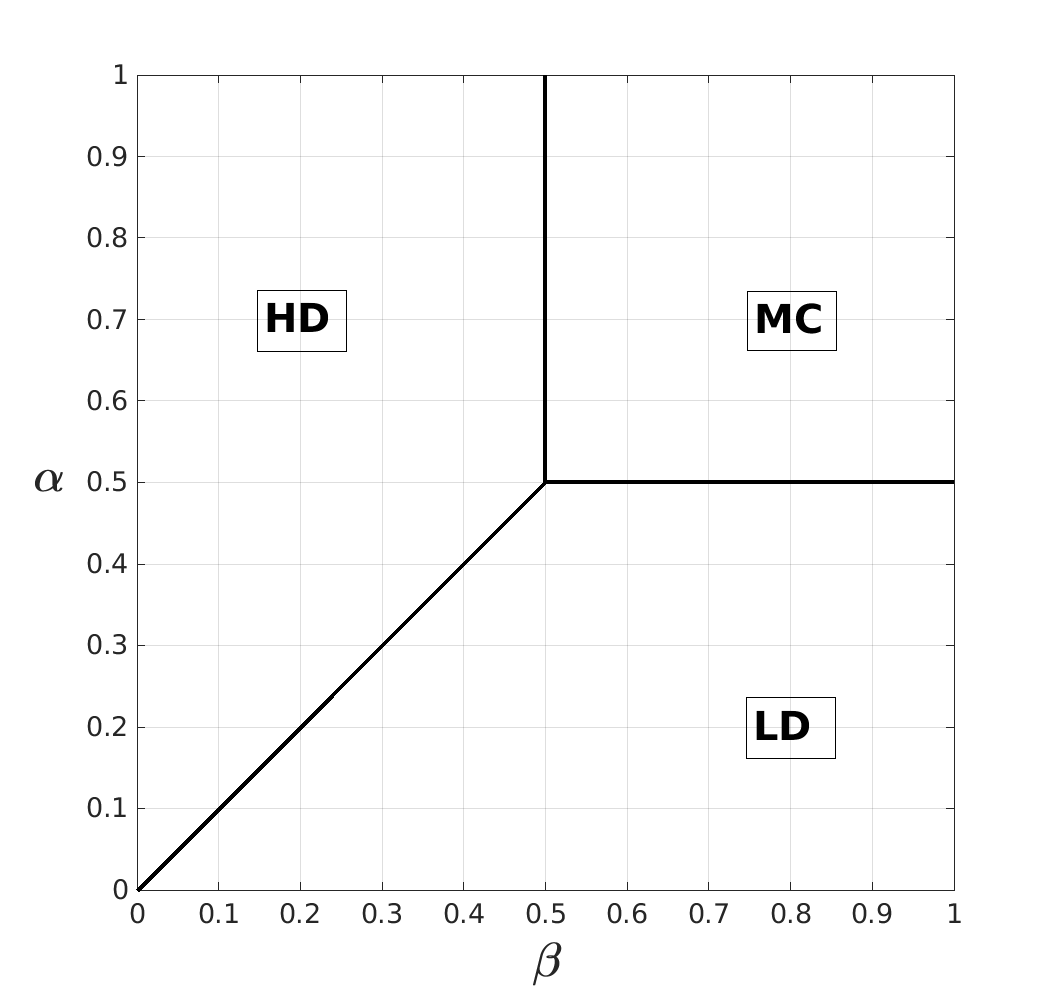}
    \caption{Phase diagram with all internal rates $h_k=1$.}
    \label{fig:phase_diagram}
\end{figure}

Our first set of numerical studies focuses on TASEP with all internal transition rates set to $h_k=1$ and leaving the entry rate $\alpha$ and the exit rate $\beta$ as parameters. For such models explicit formulas for the equilibrium of the master equation were presented in the seminal paper~\citep{Derrida1993-kr}.
They allow to compute the equilibrium values for the expected occupations $\langle \tau_k \rangle$ without solving the master equation and this was used to rigorously justify the phase diagram discovered earlier in~\citep{Krug1991}. Figure~\ref{fig:phase_diagram} displays the three different phases, called low-density (LD), high-density (HD), and maximum current~(MC) phase. We refer to the diagonal separation line between HD  and LD phase as the critical line.

\begin{figure}[htb]
    \centering
    \includegraphics[scale=0.3]{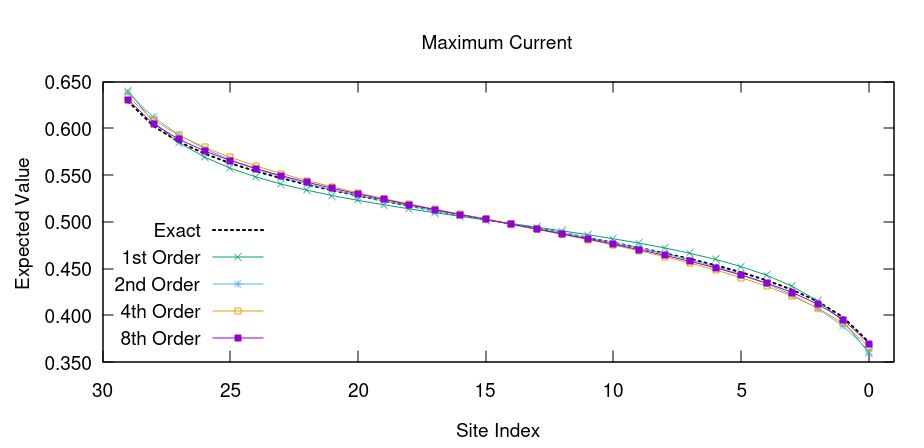}\\[-1mm]
    \includegraphics[scale=0.3]{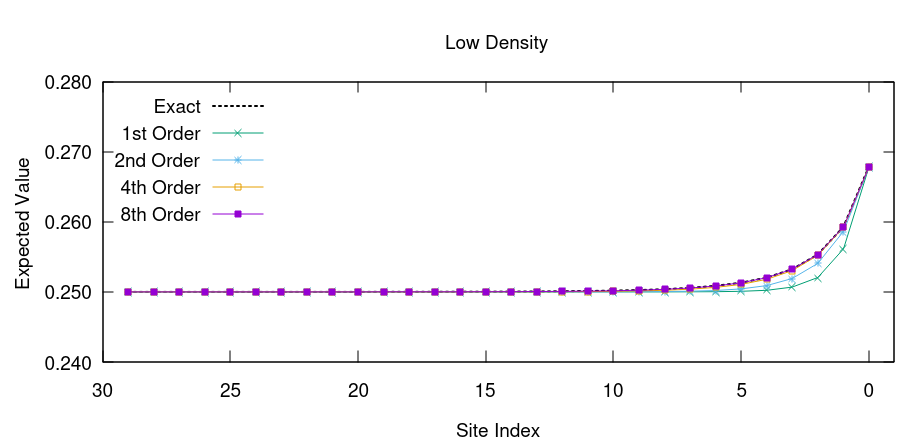}\\[-1mm]
    \includegraphics[scale=0.3]{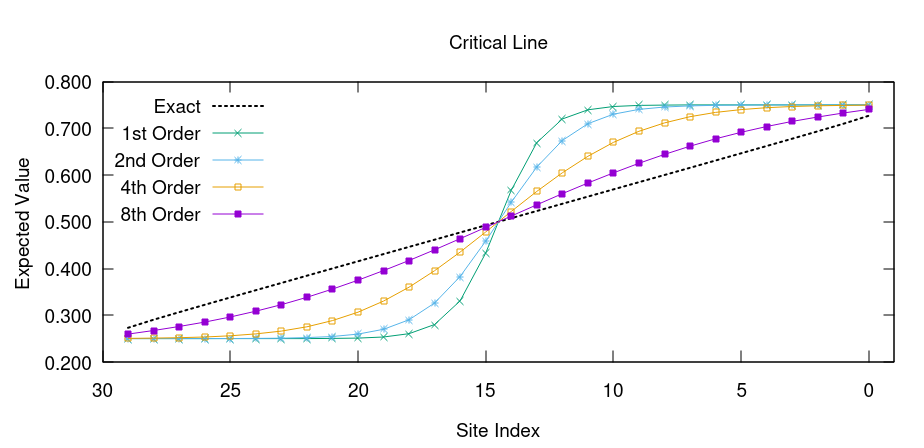}
    \caption{Equilibrium density profiles for three points in the phase diagram for lattice length~$n=30$; top to bottom: $\alpha = 0.7$, $\beta = 0.7$ (MC); $\alpha = 0.25$, $\beta = 0.7$ (LD); $\alpha = 0.25$, $\beta = 0.25$ (critical line).}
    \label{fig:LD_MC_CL_density}
\end{figure}

Figure~\ref{fig:LD_MC_CL_density} shows the equilibrium values for the expected occupations for three different points in the phase diagram.
The representative from the MC phase shows very small deviations between the exact values and all mean field approximations.
For the LD phase significant deviations appear only near the exit site and only for approximations of order $m=1$ and $m=2$. However, at the critical line the deviation of the first order model from the exact solution is very pronounced. Here we see clearly that approximations of higher order provide better results. Note that the HD phase is not included because a hole-particle duality allows to map HD to LD by interchanging the roles of $\alpha$ and $\beta$.

\begin{figure}[htb]
    \centering
    \includegraphics[scale=0.25]{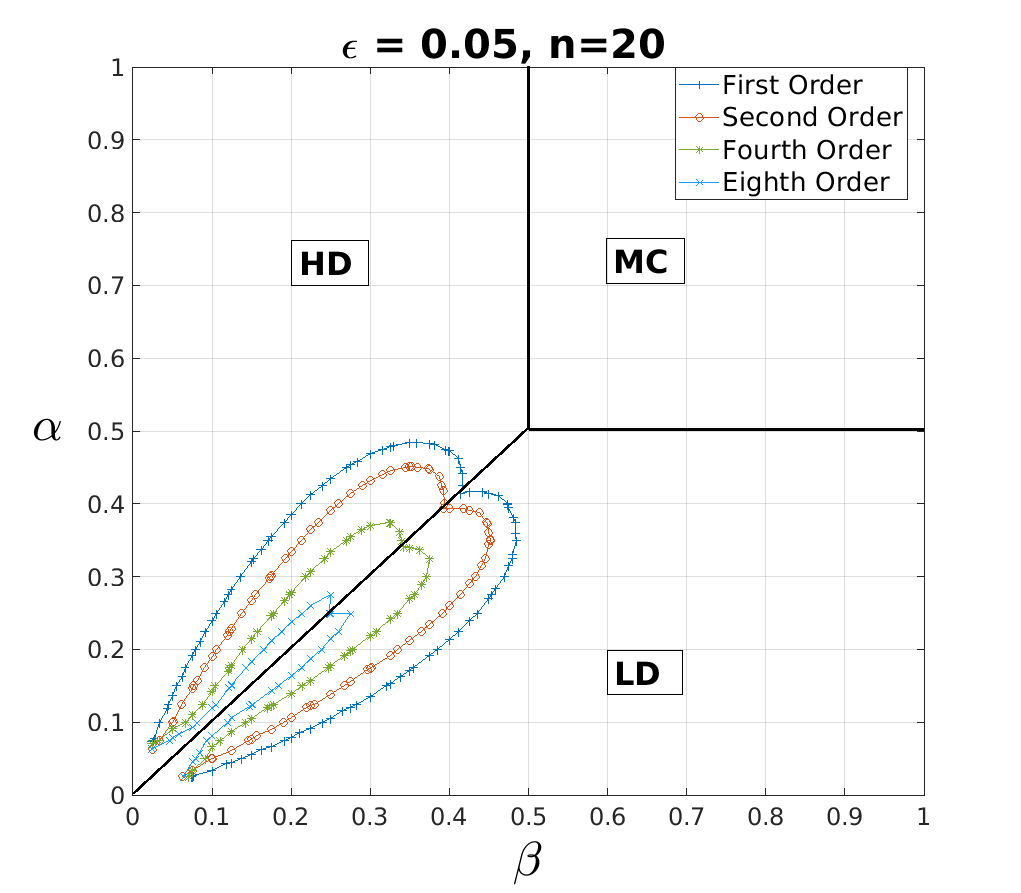}\\[-6mm]
    \caption{Contour of $\ell^2_n$-deviations, $n=20$,  $\epsilon = 0.05$
    }
    \label{fig:contour20}
\end{figure}

\begin{figure}[htb]
    \centering
    \includegraphics[scale=0.25]{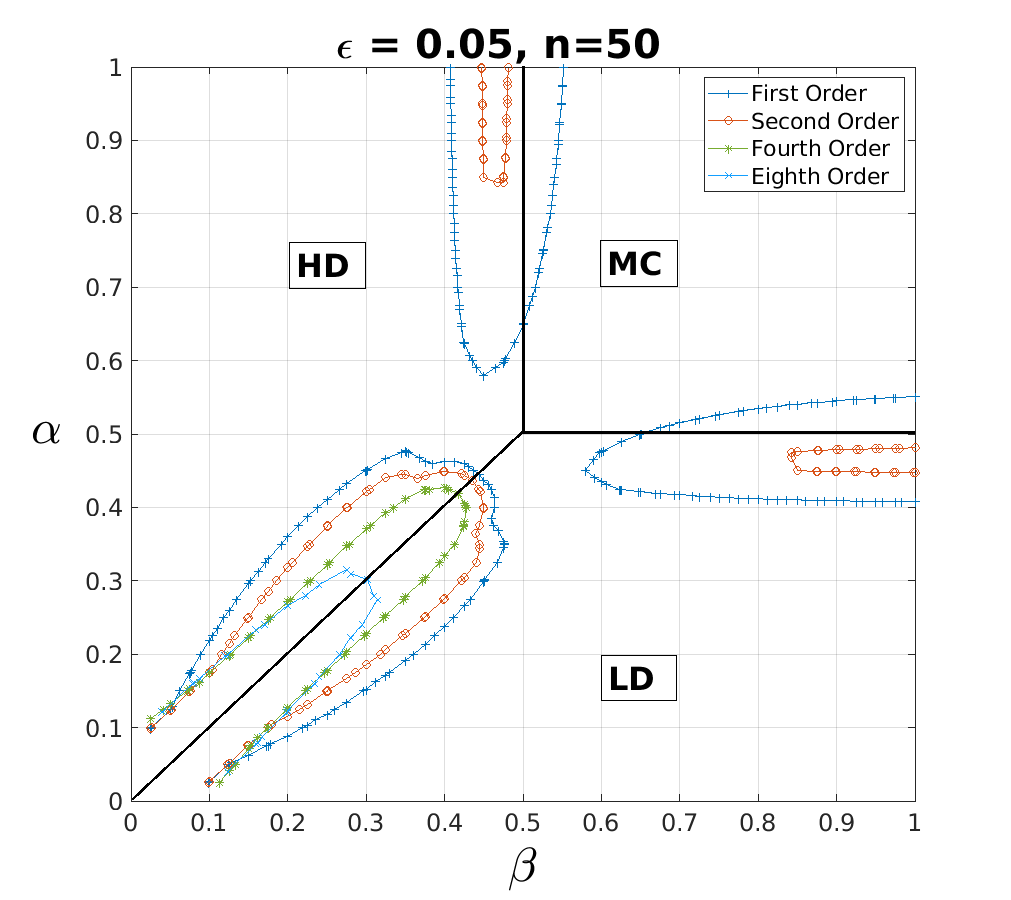}\\[-2mm]
    \caption{Contour of $\ell^2_n$-deviations, $n=50$,  $\epsilon = 0.05$
    }
    \label{fig:contour50}
\end{figure}

A more thorough examination of the entire phase diagram is shown in Figures~\ref{fig:contour20} and~\ref{fig:contour50}.
For lattice lengths $n=20$ and $n=50$, respectively, the errors were computed for 1600 points $(\alpha_i, \beta_j)=\frac{1}{40}(i,j)$ with $1\leq i, j \leq 40$. For each of these points we determined the deviations~$\epsilon$ of exact solutions and approximations (of orders $m=1,2,4,8$) by considering the density profiles as vectors in~$\mathbb{R}^n$ and measuring distances using the normalized $\ell^2_n$-norm $\|a\|_{2,n} := \sqrt{\frac{1}{n} \sum_{i=1}^n a_i^2}$. In the plots the error level $\epsilon = 0.05$ is displayed. For $n=20$ such errors only occur near the critical line. The regions with errors $\epsilon > 0.05$  lie in the interior of the corresponding level sets and they shrink as the order of the approximation increases. However, for small values of $\alpha$ and $\beta$ the error remains high, even for the approximation of order $m=8$.

For lattice length~$n=50$ deviations~$> 0.05$ appear in addition near the transitions to the MC phase, but only for approximation orders~$m=1$ and~$m=2$. As in the case $n=20$ and consistent with Figure~\ref{fig:LD_MC_CL_density}, deviations remain small away from the regions near the phase transitions. 

\begin{figure}
    \centering
    \includegraphics[scale=0.33]{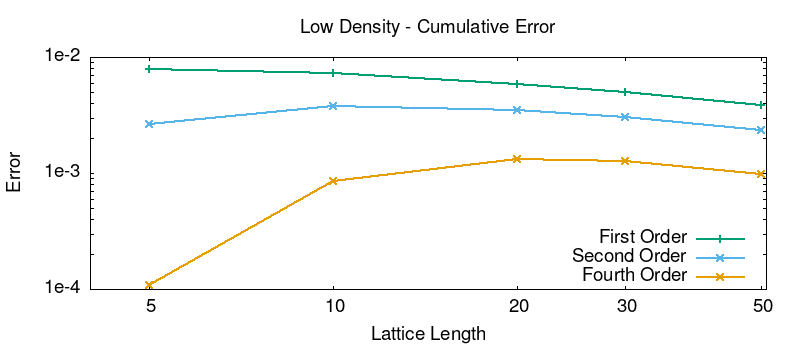}\\[-1mm]
    \includegraphics[scale=0.33]{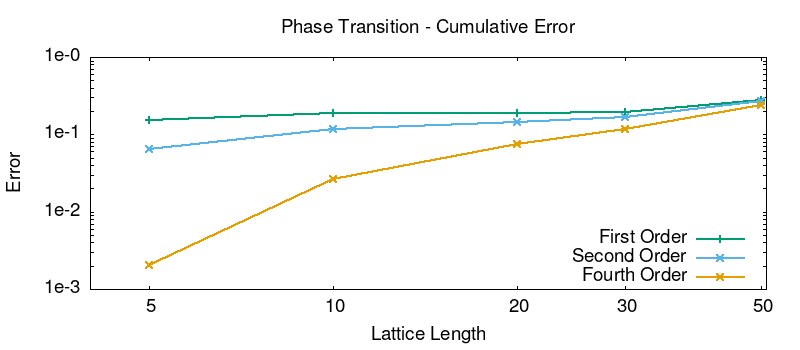}    
    \caption{Cumulative error for two parts of the phase diagram. Top figure: $\alpha \in [0.025,0.4]$, $\beta \geq \alpha+0.1$; bottom figure: $\alpha, \beta$ at most $0.1$ away from the lines of phase transition.}
    \label{fig:cumulative_phase}
\end{figure}

 Next we define two regions in the phase diagram. The LD-region consists of those points in the LD phase for which $0.025 \le \alpha \le 0.4$ and $\beta\ge \alpha + 0.1$, whereas the region of phase transitions contains the points that have an $\ell_1$-distance $\le 0.1$ from the lines of phase transitions. For each lattice length $n \in \{5, 10, 20, 30, 50\}$ and each order of approximation $m\in \{1,2,4\}$ we compute the approximation error for each region by computing its normalized $\ell^2_{qn}$-norm, where $q$ denotes the number of points $(\alpha_i, \beta_j)$ in the corresponding region. Figure~\ref{fig:cumulative_phase} shows that in the LD-region higher orders of approximation yield better results independent of the total length of the lattice. Moreover, the slight decay of the errors with growing lattice size~$n$ suggests that the deviations remain local, see also Figure~\ref{fig:LD_MC_CL_density} (middle), so that the decay is caused by the normalization of the $\ell^2_{qn}$-norm. Near phase transitions, however, the order of the approximation needs to grow with the lattice size to keep a given level of accuracy. 

\begin{figure}[htb]
    \centering
    \includegraphics[scale=0.33]{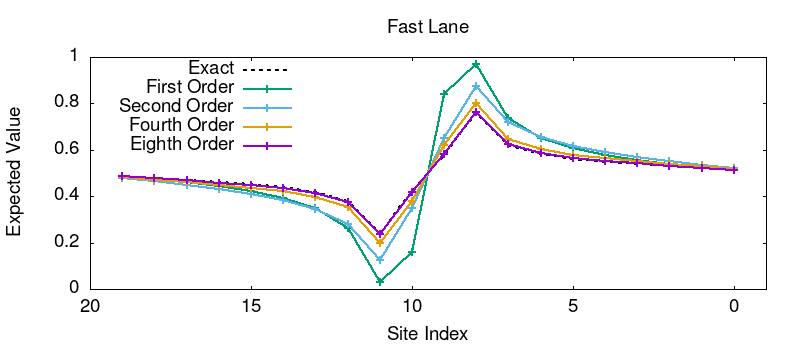}\\[-1mm]
    \includegraphics[scale=0.33]{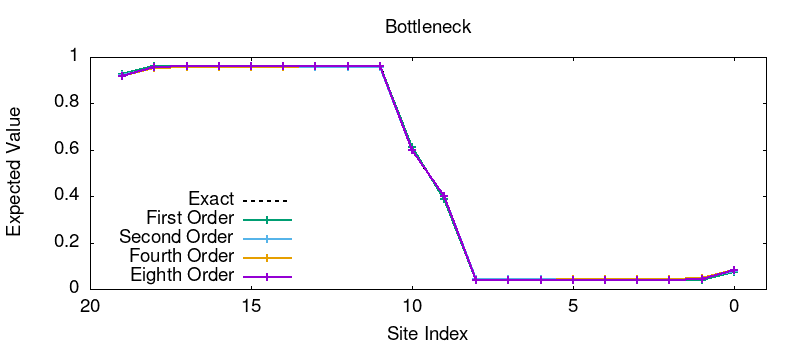}
    \caption{Equilibrium density profiles for fast lane and bottleneck.
    }
    \label{fig:fl-bl-density-profile}
\end{figure}

\begin{figure}[htb]
    \centering
    \includegraphics[scale=0.33]{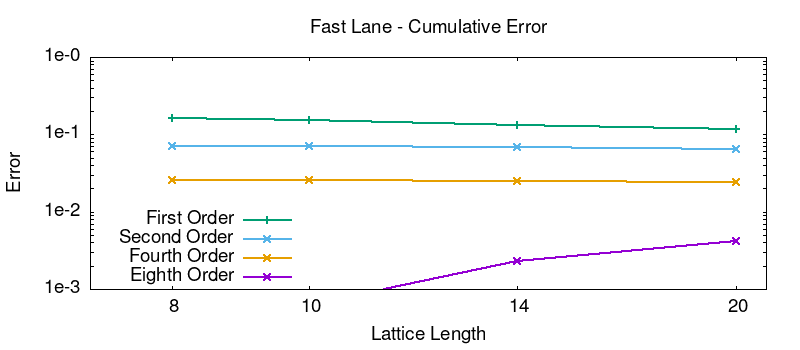}\\[-1mm]
    \includegraphics[scale=0.33]{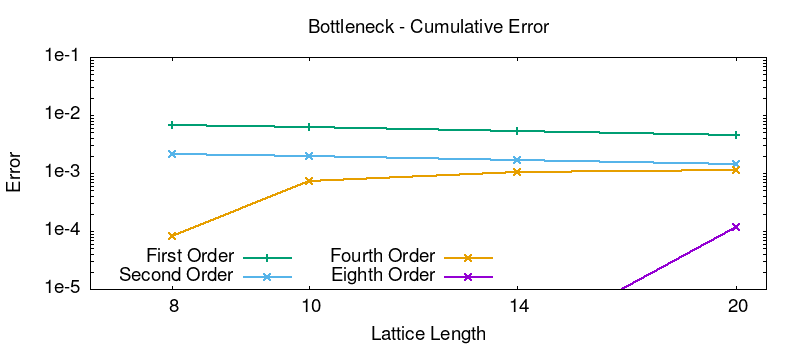}
    \caption{$\ell^2_{n}$-deviations for fast lane and bottleneck.
    }
    \label{fig:fl-bn-cumulative}
\end{figure}

In our next and final set of examples not all internal transition rates are kept equal. Using the notation introduced in Section~\ref{Sec:3} we study ``fast lanes'' and ``bottle\-necks''. More precisely, the transition rates are given by $\alpha = \beta = 0.5$ and $h_i = 1$ for all $i$, except for three consecutive sites $k$ in the middle that all have the rates $h_k=10$ (fast lane) or $h_k=0.1$ (bottleneck). Figure~\ref{fig:fl-bl-density-profile} displays the corresponding equilibrium density profiles for lattice length~$n=20$ together with approximation orders $m =1,2,4, 8$. 
Figure~\ref{fig:fl-bn-cumulative} shows the corresponding errors measured in the $\ell^2_{n}$-norm as a function of lattice length $n\in \{8,10,14, 20\}$.
Observe, that higher orders $m$ always lead to better approximations. Moreover,
the slight decay of the deviations with growing values of~$n$ again suggests that approximation errors are of a local nature. This locality is also visible in the density profiles. For the fast lane, the behavior is consistent with the analysis in the second case study in Section \ref{Sec:3}. 
For fast lane and bottleneck alike the internal transition rates $h_k$ are piecewise constant. It is striking that for the fast lane the middle part $11\geq k \geq 8$ resembles the densities at the critical line, cf.~Figure~\ref{fig:LD_MC_CL_density} (bottom), and, with some imagination, for the bottleneck it is similar to the behavior in the MC phase. In the wings we see LD/HD behavior. This observation is food for hope that the analysis of the approximation errors for constant transition rates $h_k=1$ will also shed light on the case of piecewise constant transition rates.

\section{Conclusion}\label{Sec:5}

Our theoretical error analysis in Corollary \ref{cor:approx} has shown that the error is determined by two factors, of which the term involving the conditional expectations of~$X_d$ and~$X_{d+m}$ is the faster decaying one in our case studies. Our numerical evaluations reveal that the RFM, i.e., the first order mean-field approximation, yields small errors in the MC phase and for lattices with bottlenecks. Conversely, RFM and other lower order mean-field approximations have larger errors in the LD/HD phase and in fast lane configurations. On the critical line, the order $m$ of the mean-field model needed to maintain a certain accuracy depends on the chain length, while for fast lane configurations the required~$m$ is determined by the length of the fast lane rather than by the length of the whole chain. We expect the results from this paper to provide a sound basis to make such statements rigorous in future research. If $m$ can be chosen independent of $n$, then our proposed approximation is particularly powerful, because the number of equations in the approximate model is bounded by $n2^{m+1}$ and hence only grows linearly in $n$. 

\begin{ack}
We thank Peter Koltai for useful remarks concerning pair-approximations and conditional correlations.
\end{ack}

\bibliography{main}
\end{document}